\documentclass{JHEP3}
\usepackage{amsmath}
\usepackage{epsfig}

\newcommand{\Q}{{\cal Q}}

\newcommand{\id} {{\cal I}}

\newcommand{\cO} {{\cal O}}

\newcommand{\vev}[1] {\left<#1\right>}

\title{Comments on superstring field theory and its vacuum solution}

\author{Michael Kroyter\\ \\
Center for Theoretical Physics\\
Massachusetts Institute of Technology\\
Cambridge, MA 02139, USA\\
\\ and\\ \\
School of Physics and Astronomy\\
The Raymond and Beverly Sackler Faculty of Exact Sciences\\
Tel Aviv University, Ramat Aviv, 69978, Israel\\ \\
\email{mikroyt@mit.edu}, \email{mikroyt@tau.ac.il}}

\abstract{We prove that the NS cubic superstring field theories are
classically equivalent, regardless of the choice of $Y_{-2}$ in their
definition, and illustrate it by an explicit evaluation of the action
of Erler's solution. We then turn to examine this solution. First, we explain
that its cohomology is trivial also in the Ramond sector. Then, we show that
the boundary state corresponding to it is identically zero.
We conclude that this solution
is indeed a closed string vacuum solution despite the absence of a tachyon
field on the BPS D-brane.}

\preprint{MIT-CTP-4036\\TAUP-2899-09}

\begin{document}

\section{Introduction}

In this paper we address two issues. First, we deal with the definition of
the NS sector of cubic superstring field theory. The action of this theory
is given by~\cite{Preitschopf:1989fc,Arefeva:1989cm,Arefeva:1989cp},
\begin{equation}
\label{action}
S=-\int Y_{-2} \Big(\frac{1}{2}\Psi Q\Psi+\frac{1}{3}\Psi^3\Big),
\end{equation}
where $\Psi$ is a ghost-number one, picture-number zero string field and
$Y_{-2}$ is a mid-point insertion of the double inverse picture changing
operator.
From this action one derives the equation of motion,
\begin{equation}
\label{EOM}
Q\Psi+\Psi^2=0\,.
\end{equation}
There are several reasons for criticizing this theory, the most important
of which is the problem with defining the gauge transformation upon the
inclusion of the Ramond sector~\cite{Kroyter:2009zi}. Another,
``aesthetical''
objection comes from the fact that there are many ways to define the $Y_{-2}$
insertion. We confront this problem in section~\ref{sec:equiv}, where we
prove that classically all these theories are equivalent. We illustrate
the general prove by an explicit calculation of the action for the case
of Erler's solution~\cite{Erler:2007xt}.

Erler's solution is the subject of the rest of the paper.
It is a supersymmetric generalization of Schnabl's
solution~\cite{Schnabl:2005gv} of bosonic string field
theory~\cite{Witten:1986cc}. The study of Schnabl's solution
proved~\cite{Schnabl:2005gv,Okawa:2006vm,Fuchs:2006hw,Ellwood:2006ba} that
some of Sen's conjectures~\cite{Sen:1999mh,Sen:1999xm} hold in the
framework of string field theory, which proved to be adequate for describing
non-perturbative solutions\footnote{See~\cite{Fuchs:2008cc} for a review of
these and other results.}.
The idea of generalizing Schnabl's solution to the
supersymmetric theory was suggested in~\cite{Fuchs:2007gw}. There, a
solution of the non-polynomial superstring field
theory~\cite{Berkovits:1995ab} was presented.
Unfortunately, it turned out that this solution is trivial.
Then, Erler presented his solution in the context of cubic
superstring field theory.
Using the equivalence between the two formalisms of superstring field
theories~\cite{Fuchs:2008zx}, Erler's solution was mapped to a
solution, which differs from the one suggested in~\cite{Fuchs:2007gw}
only by the location of the $P$ insertion.

In split-string notations~\cite{Erler:2006hw,Erler:2006ww}, Erler's
solution is given by,
\begin{equation}
\label{ErlerSol}
\Psi=Fc\frac{KB}{1-F^2}cF+FB\gamma^2 F\equiv \Psi_S+\tilde\Psi\,,
\end{equation}
where $\Psi_S$ is Schnabl's solution and the insertion defining
$\tilde \Psi$ can be written in the fermionized
variables~\cite{Friedan:1985ge} using,
\begin{equation}
\gamma^2=\eta\partial\eta e^{2\phi}\,.
\end{equation}
Furthermore, Erler found that his solution has the correct tension for
cancelling the original D-brane tension and that the cohomology around
the solution is trivial, in accord with the interpretation of this solution
as a closed string vacuum solution.

Despite the success of Erler's solution in reproducing the physics of
the closed string vacuum, it was claimed in~\cite{Aref'eva:2008ad} that
this solution cannot represent this vacuum, since it is defined on a
BPS D-brane that does not support a tachyon field. Moreover, it was
suggested that a variant of Erler's solution that is defined only on a
non-BPS D-brane can describe the non-perturbative vacuum. It was also shown
that this variant has the correct action. The cohomology of this variant
was shown to be trivial in~\cite{Fuchs:2008zx}. However, it was also shown
there that the two solutions are a part of a one-parameter family of
solutions. All these solutions share the properties of having the correct
action and a trivial cohomology. Furthermore, it was claimed that they are
all gauge equivalent and the gauge transformations were written explicitly.
These gauge transformations seem to be regular and the contracting homotopy
of the kinetic operator around the solutions transforms trivially with
respect to them. It seems that if Erler's solution does not describe
the non-perturbative vacuum, these solutions do not describe it as well.
Hence, we should decide whether we accept that Erler's solution represents
the closed string vacuum or reject it and look for other, new solutions
(presumably defined only on non-BPS D-branes).

Of course, rejecting the natural interpretation for Erler's solution without
providing an alternative
explanation to the results regarding its cohomology and action could not be
satisfactory. Indeed, an alternative point of view on
these matters was presented in~\cite{Aref'eva:2008ad}.
According to this proposal, the cohomology of Erler's solution is trivial
only for the NS sector. Hence, the solution presumably represents
a supersymmetry breaking phase. In
section~\ref{sec:Ramond}, we explain why this interpretation is wrong.

In order to decide what is the physical meaning of Erler's solution,
we evaluate the boundary state associated with it in
section~\ref{sec:boundary}.
The boundary state associated with a given solution was constructed
by Kiermaier, Okawa and Zwiebach in~\cite{Kiermaier:2008qu}\footnote{This
relies on the construction of gauge invariant overlaps of open-closed
strings~\cite{Hashimoto:2001sm,Gaiotto:2001ji,Michishita:2004rx,
Ellwood:2008jh,Kawano:2008ry,Kawano:2008jv,Kishimoto:2008zj} as well as
on techniques devised in the evaluation of scattering amplitudes in
Schnabl's gauge~\cite{Rastelli:2007gg,Kiermaier:2007jg,Kiermaier:2008jy}.}.
The boundary state carries
a full information on the BCFT that the solution describes. Thus, its
triviality in the case of Erler's solution proves that this
solution indeed corresponds to the closed string vacuum.
It seems that string field theory does not need a tachyon in order to
be able to describe this vacuum.
We return to this issue and offer further concluding remarks
in section~\ref{sec:conc}.

\section{The classical equivalence of cubic NS superstring field theories}
\label{sec:equiv}

The cubic superstring field theory was criticized on several grounds.
The most commonly made reservation is related to the use of picture changing
operators in the definition of the action. We believe, that at least
classically, this fact by itself should not pose any
problem~\cite{Kroyter:2009zj}. However, there is a genuine problem, namely
that of using picture
changing operators in defining the gauge symmetry in the Ramond
sector~\cite{Kroyter:2009zi}. These operators collide when one tries to
iterate the linearized gauge transformation. Hence, the finite form of the
gauge transformation does not exist and the theory cannot be expected to
describe string theory. Nonetheless, one might still hope that, at the
classical level and when restricted to the NS sector, the theory still
makes sense and has some predictive power. This is indeed the case as
can be seen from the existence of Erler's solution.

In addition, an ``aesthetic'' reservation exists against the
NS sector of the cubic theory. The matter here is the
appearance of the $Y_{-2}$ operator in the definition of the theory,
as this operator is not unique. Note that the space of
operators obeying all the properties that $Y_{-2}$ should have is seven
dimensional~\cite{Preitschopf:1989fc}. Of this space, only a one dimensional
sub-space is left after identifying operators that differ by $Q$ exact terms.
Nonetheless, while this resolves the ambiguity for the worldsheet theory,
it does not resolve it a-priori within string field theory. Moreover, the
fact that we have two ``mid-points'' in our disposal, namely $\pm i$,
implies that one can use an arbitrary linear combination of terms, such
that each term is the product of two local picture changing operators,
whose total picture number is $-2$,
\begin{equation}
Y_{-2}=\sum_n u_n X_n(i) X_{-2-n}(-i)\,.
\end{equation}
Here, the $u_n$ are coefficients and $X_n$ is defined as a picture changing
operator that changes the picture by $n$ units.
In particular, $X_{-2}$ is (a specific local choice of) $Y_{-2}$.
The freedom of adding exact terms in the
definition of the $X_n$'s remains. This construction
implies that the space of superstring field theories defined is in fact
infinite dimensional and it is not a-priori clear whether they are all
equivalent and if not, which one is the correct one\footnote{There is an
important restriction on the form of $Y_{-2}$: It has to be a
primary conformal field. We return to this issue below.}.

It is usually claimed that the ``non-chiral'' theory obtained by defining
\begin{equation}
\label{nonChirYm2}
Y_{-2}=Y(i)Y(-i)\,,
\end{equation}
is the correct one, since ``the other theory'' does not obey
twist symmetry. However, as we stressed, there are infinitely many ``other
theories''.
Many of these theories do obey twist symmetry. Thus, the
only obvious reason to prefer the theory~(\ref{nonChirYm2}) is its
simplicity. This is not a strong enough argument when it comes by itself.
A stronger argument in favour of the ``non-chiral'' theory is the recently
established equivalence~\cite{Fuchs:2008zx,Kroyter:2009zi} between it and the
non-polynomial theory~\cite{Berkovits:1995ab,Berkovits:2001im,
Michishita:2004by}\footnote{This equivalence is classical, and is defined up
to issues of regularization of a mid-point insertion in its definition. It
is then defined in the NS sector, although formally it holds also in the
Ramond sector, despite the problems with the definition of the gauge
symmetry in this case.}.

One might still not be happy about the choice of the chiral insertion,
since it seems that it is fixed not by its own merits, but by an
equivalence to another, more established, formalism. Also, the
proof of equivalence does not rule out the possibility that some of the
other theories are also equivalent to the non-polynomial one.
Here, we want to show that all these theories are classically
equivalent, i.e., they all have the same solutions and gauge symmetry, they
define the same boundary states and the actions of these solutions do not
depend on the specific choice of $Y_{-2}$.
The proof that the solutions and gauge transformations are the same
follows from assuming that the space
of string fields is defined in a way that avoids potential zeros
and singularities with all the $Y_{-2}$'s. The simplest possibility
is to assume that the space of string fields contains no states
with mid-point insertions~\cite{Kroyter:2009zj}.
The assertion regarding the boundary state is then trivial, since it depends
only on the solution itself.
What is left to prove then, is only that given a solution, its action is the
same regardless of the choice of $Y_{-2}$.
We prove this assertion in~\ref{sec:ProofGen}\footnote{Our proof is
reminiscent of the analysis at the end of section 4
of~\cite{Witten:1986qs}.}.
Then, in~\ref{sec:ProofErler},
we evaluate the action of Erler's solution in the theory with a chiral
$Y_{-2}$ insertion and show explicitly that it is the same as in the case
of a non-chiral insertion evaluated in~\cite{Erler:2007xt}.

\subsection{The general proof}
\label{sec:ProofGen}

Let there be two theories that differ by their $Y_{-2}$ insertion. Then, up
to the total scaling that should be canonically fixed, these insertions
differ by a $Q$-exact term in their definition,
\begin{equation}
\label{QYY}
Y_{-2}^{(2)}-Y_{-2}^{(1)}=Q\Upsilon\,.
\end{equation}
The case where both mid-points are used also falls under this
definition. The reason being that the local $Y_{-2}$ can be defined as,
\begin{equation}
\label{Y2YY}
Y_{-2}(w)=\oint_w\frac{dz}{2\pi i}\frac{Y(z)Y(w)}{z-w}\,,
\end{equation}
and similarly for the other picture changing operators.
Adding a $Q$-exact term $Q \Xi(i)$ to the insertion at $z=i$, while not
changing that at $z=-i$, can be achieved by considering,
\begin{equation}
\delta X_n(i) X_{-2-n}(-i)=Q\big(\Xi_n(i)X(-i)\big).	
\end{equation}
Moving $X$ from one point to another can be achieved by the small Hilbert
space exact term,
\begin{equation}
X(i)-X(-i) =Q\big(\xi(i)-\xi(-i)\big)\,.
\end{equation}
Similarly, $Y$ can be moved since it is also exact in the large Hilbert
space,
\begin{equation}
Y(z)=Q\Big(\frac{i}{5}\,c\,\xi\partial\xi e^{-3\phi}\psi\cdot \partial X
  -\xi e^{-2\phi}\Big)\,.
\end{equation}
Expressions like the ones above can be used in order to turn an insertion
of the form $X_n X_{-n-2}$ into an insertion of the form
$X_{n\pm 1} X_{-n-2\mp 1}$.
Hence, what we have to show is that two theories, whose mid-point insertions
differ as in~(\ref{QYY}), are classically equivalent.

Let us now write the difference in the action~(\ref{action}) of a given
solution, between two theories that differ as in~(\ref{QYY}),
\begin{equation}
\label{genProof}
\delta S=\frac{1}{6}\int \Psi^3 Q\Upsilon=
\frac{1}{2}\int Q\Psi\Psi^2 \Upsilon=
-\frac{1}{2}\int\Psi^4 \Upsilon=\frac{1}{2}\int\Psi^4 \Upsilon=0\,.
\end{equation}
Here, in the first equality, the equation of motion~(\ref{EOM}) was used.
Then, we integrated $Q$ by parts and used the fact that $\Upsilon$
is an odd mid-point insertion in order to rearrange the various
terms. In the third equality, we used
the equations of motion again. Next, we used once more the cyclicity of the
integral and the fact that $\Upsilon$ is a mid-point insertion in order to
move the first $\Psi$ to the last position, picking a minus sign on the way.
This implies that the expression vanishes and the proof is complete.

There are two potential difficulties with the proof above:
\begin{itemize}
\item The proof~(\ref{genProof}) assumes a local mid-point insertion.
      However,~(\ref{Y2YY}) uses a neighbourhood of the mid-point. We
      believe that this is not really a problem, since the contour can be
      made arbitrarily small and hence the manipulations of~(\ref{genProof})
      can be justified up to an arbitrary accuracy, for an arbitrary
      solution that carries no mid-point insertions.
\item The $Y_{-2}$ insertions have to be primary conformal fields.
      Nonetheless, it might
      seem that the proof works regardless of this requirement. Indeed,
      one may consider a particular conformal frame for the evaluation of the
      action, in which changing the order of the fields is described by an
      SL(2) transformation. An example of such a frame is the unit disk,
      cut into equal wedges. Changing $Y_{-2}$ to a weight zero
      non-primary insertion works fine in this coordinates. However, if we
      want the theory to be well defined, regardless of a conformal frame,
      we should insist on having insertions of (zero weight) primaries at
      both mid-points.
      Then, the proof above works, provided that $Q\Upsilon$ is primary,
      i.e., provided we are relating two legitimate theories.
\end{itemize}

\subsection{An explicit calculation: Erler's solution}
\label{sec:ProofErler}

In the developments following Schnabl's solution, analytical solutions were
constructed that describe vacuum solutions and marginal
deformations~\cite{Schnabl:2007az,Kiermaier:2007ba,Erler:2007rh,Okawa:2007ri,
Okawa:2007it,Fuchs:2007yy,Kishimoto:2007bb,
Fuchs:2007gw,Kiermaier:2007vu,Kiermaier:2007ki}.
The marginal deformations depend continuously on a parameter. The
derivative of the action with respect to this parameter gives an integrand
that is proportional to the equation of motion. Hence, the action of the
marginal solutions is zero, as is adequate for a solution that describes a
marginal deformation. Thus, we cannot use these solutions for a non-trivial
verification of the proof above. The only other analytical solution at our
disposal is Erler's solution (and its gauge equivalent
ones~\cite{Aref'eva:2008ad,Fuchs:2008zx}).

In~\cite{Erler:2007xt}, Erler evaluated the action of his solution using the
bi-local version of $Y_{-2}$~(\ref{nonChirYm2}). According to our discussion,
the same value for the action should be obtained upon evaluating the
action of this solution using a local primary insertion. As we already
stated, there is (up to a scaling) a seven dimensional space of potential
$Y_{-2}$'s.
However, not all of them are primary. We consider a particular, primary
representative in this space and normalize it canonically, i.e., we demand
that it obeys the OPE,
\begin{equation}
Y_{-2}(z)X(w)= Y(w)+\cO(z-w)\,.	
\end{equation}
The insertion we consider was presented already in~\cite{Arefeva:1989cp}.
It is given by,
\begin{equation}
\label{ChirY2}
Y_{-2}(z)=-e^{-2\phi(z)}
   -\frac{i}{5}c \partial \xi e^{-3\phi}\psi_\mu \partial X^\mu (z)\,.
\end{equation}

For a solution, the action~(\ref{action}) can be reduced to,
\begin{equation}
S=\frac{1}{6}\int Y_{-2} \Psi^3\,.
\end{equation}
The solution~(\ref{ErlerSol}) has no explicit dependence on the matter
($X^\mu$ and $\psi^\mu$) sectors\footnote{It has implicit ones, since $K$ is
an integral of the total energy momentum tensor. Moreover, the wedge states
that appear in the expansion also depend on $K$. However, this dependence
has a geometrical interpretation in terms of surfaces on which the
expectation value should be evaluated. Hence, our conclusions do not
change.}. Thus, the second term in~(\ref{ChirY2}) cannot contribute.
Inspecting the solution~(\ref{ErlerSol}) further, we see that the second
term of this solution cannot contribute, since a total $\phi$ charge of
$-2$ is necessary. This charge is exactly supplied by the first term
of~(\ref{ChirY2}) and there is no term that can decrease it. Hence, terms
that increase it will not contribute to the action.

We conclude that we are left with,
\begin{equation}
S=\frac{1}{6}\int e^{-2\phi} \Psi_S^3\,,
\end{equation}
where $\Psi_S$ is the first term in~(\ref{ErlerSol}). Now, the only
explicit $\phi$ dependence appears in the insertion. The $\Psi_S^3$
implies that this term should be evaluated on various wedges and
that derivatives with respect to wedge size should be performed,
but only after evaluating the expectation value in all sectors.
The $e^{-2\phi}$ insertion is a weight zero primary. Hence, its
expectation value is surface-independent and equals one in the conventions
we use here. Evaluating the trivial $\phi$-sector expectation value
leaves us with,
\begin{equation}
S=\frac{1}{6}\vev{\Psi_S^3}\,,
\end{equation}
where the expectation value now is only in the $bc$ sector.
However, this is exactly the expression for the action of Schnabl's
solution, which is equal to Erler's one.
It is interesting to note that the evaluation of the action of Schnabl's
solution~\cite{Schnabl:2005gv,Okawa:2006vm,Fuchs:2006hw} is technically
very different from the one used by Erler for the case of the non-chiral
$Y_{-2}$~\cite{Erler:2007xt}.
Hence, the evaluation performed here gives a non-trivial verification of
the general case proved above.

Our choice of a chiral $Y_{-2}$ was criticized for not being twist invariant
as well as for some peculiar properties of its level
expansion~\cite{Urosevic:1990as}.
We conclude, that explicit twist symmetry of the action might not be that
important, at least classically, and that the strange low-level behaviour
found with this insertion is merely a level-truncation artifact.

\section{The triviality of the cohomology in the Ramond sector}
\label{sec:Ramond}

An alternative interpretation of the physical meaning of Erler's solution
was suggested in~\cite{Aref'eva:2008ad}, following similar proposal for
the interpretation of a level-truncated precursor of the same
solution~\cite{Arefeva:1990ei}. The interpretation is that of a
solution that breaks supersymmetry. Hence, it was suggested that while
perturbative NS degrees of freedom are absent around this solution, Ramond
degrees of freedom remain. Here, we prove that the cohomology is trivial also
in the Ramond sector.

In order to decide on this matter, the theory should be capable of describing
the Ramond sector. While we claimed in~\cite{Kroyter:2009zi} that the
Ramond sector is not well-described by the cubic theory, it is well-described
at the linearized level. Hence, we believe that we can decide on this matter
from studying the Ramond sector of this theory. Alternatively, we may say
that our conclusion on this matter are founded to the same degree in which
the question is well defined.

Stating the above, the proof is identical to the proof
in the NS sector. The absence of perturbative modes for the theory expanded
around the vacuum solution was demonstrated by defining a ghost-number $-1$
state $A_{\id}$ satisfying
\begin{equation}
\Q A_\id=\id\,,
\end{equation}
where $\Q$ is the kinetic operator around the solution and
$\id$ is the identity string field.
The form of $A_\id$ was found in~\cite{Ellwood:2006ba} for the case
of Schnabl's solution. Then, it was shown
in~\cite{Erler:2007xt} that the same string field works also for the
vacuum solution of the cubic superstring field theory. For the
generalizations of Erler's solution, introduced
in~\cite{Aref'eva:2008ad,Fuchs:2008zx}, it was shown in~\cite{Fuchs:2008zx},
that again the same $A_\id$ is still adequate. It was also shown there that
these generalizations are in fact gauge equivalent to Erler's solution.

All that is needed now for proving the triviality in the Ramond sector
of the solutions of~\cite{Erler:2007xt,Aref'eva:2008ad,Fuchs:2008zx},
is to note that the same kinetic operator is used in this sector and in
the NS sector. This fact is blurred in the operator representation,
where the expansion of $Q$ is in terms of different oscillator modes.
Nonetheless, the conformal current $J_B$ defining $Q$ is the same in
both cases and in terms of conformal fields the Ramond property of the
string field is expressed by the use of spin field in its
definition~\cite{Friedan:1985ge}\footnote{This argument is close in spirit
to the arguments in open-closed string field theory~\cite{Zwiebach:1997fe}
where $Q$ is moved from an open string field to a closed string field or
vice versa.}.
Thus, we conclude that the Ramond sector cohomology is also trivial, as
stated.

\section{The boundary state of Erler's solution}
\label{sec:boundary}

We would like to understand the physics behind Erler's solution.
The question arises: Which objects can we extract from the form of the
solution that would characterize its physical meaning?
Two obvious entities are the action of the solution and the cohomology
around it.
Another important example is the boundary state defined by the solution.
A boundary state is equivalent to a BCFT. Hence, defining a boundary state
using a classical solution holds a lot of information regarding its
properties.

Following~\cite{Hashimoto:2001sm,Gaiotto:2001ji}, it was shown by
Ellwood~\cite{Ellwood:2008jh} that a coupling of a closed string to
a classical solution gives information, which is related to the boundary
state of the new BCFT. However, only on-shell closed string states are
allowed in this construction, despite the fact that the boundary state
is not restricted to this case. The origin of this restriction is the
use of the string mid-point for the insertion of the vertex operator
defining the closed string state. While the mid-point is the only point
invariant under the star product~\cite{Witten:1986qs}, it is also infinitely
rescaled upon the contraction with the identity string field used in the
constructions of~\cite{Hashimoto:2001sm,Gaiotto:2001ji,Ellwood:2008jh}.
The only operators that can be consistent with such an infinite rescaling
are the scalars, i.e., the primary zero-weight conformal field that
describe on-shell closed strings.

A way around this difficulty was devised in~\cite{Kiermaier:2008qu},
where the conical singularity was replaced by a closed string local
coordinate patch of arbitrary size. In this way, the boundary state
itself can be defined (up to a possible gauge transformation) in terms of
the classical solution. This construction makes the identification of
the closed string vacuum extremely simple, since the boundary state
that corresponds to this vacuum vanishes identically.
Indeed, in~\cite{Kiermaier:2008qu}, it was shown that this construction
gives an identically zero boundary state (after a non-trivial calculation)
for Schnabl's solution.
Here, we show that the boundary state defined by Erler's solution
also vanishes identically.

Our first task is to define the boundary state in the case of superstring
field theory, since~\cite{Kiermaier:2008qu} dealt with the bosonic case.
Ideas regarding the needed generalization were proposed
in~\cite{Michishita:2004rx,Ellwood:2008jh,Kiermaier:2008qu}.
All these papers dealt with the non-polynomial
theory~\cite{Berkovits:1995ab,Berkovits:2001im,Michishita:2004by}.
However, the case of the cubic superstring field
theory~\cite{Preitschopf:1989fc,Arefeva:1989cm,Arefeva:1989cp}, is
even simpler. The relation between the constructions in the cases
of the cubic and non-polynomial theories can be understood in terms
of the classical equivalence between these
formulations~\cite{Fuchs:2008zx,Kroyter:2009zi}.
Let us describe these suggestions.

For the one-open-one-closed gauge invariant string vertex,
Michishita~\cite{Michishita:2004rx} proposed to use,
\begin{equation}
\label{Michishita}
\oint V \Phi\,.
\end{equation}
Here, $\Phi$ is the open string field of the non-polynomial theory,
$V$ is the closed string vertex operator, which is inserted at the string
mid-point and $\oint$ represents the evaluation of the expectation value
in the large Hilbert space, in which $\Phi$ resides.
To get the analogous expression in the cubic theory we assume that the
string field $\Phi$ is related to the cubic string field $\Psi$
by~\cite{Fuchs:2008zx},
\begin{equation}
\label{Map}
\Phi=P \Psi\,,
\end{equation}
where,
\begin{equation}
\label{Pdef}
P=\xi Y=-c\xi\partial \xi e^{-2\phi}\,,
\end{equation}
which is the contracting homotopy operator for $Q$ in the large Hilbert
space, is inserted at the mid-point.
The only presence of the $\xi$ zero mode in~(\ref{Michishita}) comes from
the $\xi$ in the definition of $P$~(\ref{Pdef}). Thus,~(\ref{Michishita})
can be written as,
\begin{equation}
\int Y V \Phi\,.
\end{equation}
Now, $Y$ and $V$ are inserted at the mid-point, i.e., at $\pm i$ in the
standard coordinates. Since $V$ is on-shell, acting on it with $Y$ is a
legitimate picture changing and the resulting expression is just the
natural coupling of open and closed strings,
\begin{equation}
\int V \Phi\,,
\end{equation}
where now $V$ is written in the correct picture to begin with.
A variant of this construction was suggested by
Ellwood~\cite{Ellwood:2008jh}. There, it was not assumed that the closed
string $V$ obeys,
\begin{equation}
QV=0\,,
\end{equation}
while an explicit $Q$ was assumed to act on $\Phi$.
Integrating by parts leads to the same expression as before, only with
the restriction that $V$ is not only closed, but is exact.
The motivation for this change was the relation between the non-polynomial
and the cubic theories,
\begin{equation}
\label{invMap}
\Psi=e^{-\Phi} Q e^\Phi\,,
\end{equation}
since, upon integration, the r.h.s reduces to $Q\Phi$.
Note, that~(\ref{invMap}) is exactly the inverse mapping used
in~\cite{Fuchs:2008zx}.

Indeed, in~\cite{Kiermaier:2008qu}, it was suggested that in light of the
above, the generalization to the non-polynomial theory of their boundary
state can be achieved by replacing everywhere $\Psi$ by $e^{-\Phi} Q e^\Phi$.
They also suggested that this construction can be used, as we are doing here,
in order to decide whether the vacuum solution of~\cite{Fuchs:2008zx} is
indeed a vacuum solution. This solution is, however, nothing but a mapping
under~(\ref{Map}) of Erler's solution. Hence, using the inverse
map~(\ref{invMap}), we conclude that the suggestion
of~\cite{Kiermaier:2008qu} corresponds to using their formalism
without any modification in the cubic superstring field theory.
This gives gauge covariant expressions by construction, since the
algebraic properties of the bosonic theory and the cubic superstring
field theory are the same. Then, obtaining a trivial boundary state for
Erler's solution implies that Erler's solution and its analogous
one in the non-polynomial theory, are closed string vacuum solutions.

The boundary state has a very clear geometric representation
(see~\cite{Kiermaier:2008qu} for more details).
One takes a propagator at some specific gauge and cuts it in half along the
trajectory of the string mid-point. There exists a natural coordinate system
in which the cut line and the original boundary are both horizontal and
the whole half-propagator strip is obtained by horizontal translation, as in
fig.~\ref{fig:halfProp}. The left and right curves are identified.
Now, cuts, whose forms are also obtained by
horizontal translations are introduced. The number of cuts is summed over
and their location is integrated over. Into each cut one has to glue a
factor of $[B_R,\Psi]$, where $B_R$ is a specific $b$-ghost line integral.
For the case of Schnabl's solution it was shown that
this construction gives a vanishing boundary state. Erler's solution contains
two pieces~(\ref{ErlerSol}). The first term is identical to Schnabl's
solution from a geometrical
point of view\footnote{The Virasoro generators used now are the ones of the
NS theory. Hence, strictly speaking the first part of Erler's solution
differs from Schnabl's solution. Nonetheless, the results of all the
evaluations depends on the induced geometry. Thus, this part gives exactly
the same result in the NS theory, as was obtained from Schnabl's solution
in the bosonic theory.}.
\FIGURE{
\label{fig:halfProp}
\epsfig{figure=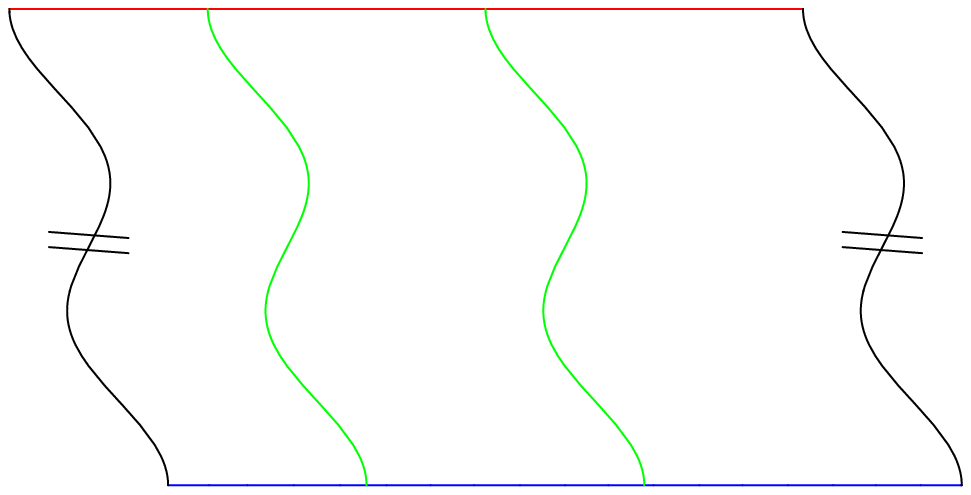, width=9cm}
\caption{Constructing the boundary state for a generic half-propagator:
The left and right (black) curves
are identified. The bottom (blue) line has the boundary conditions of the
original BCFT. The top (red) line gives (after the identification) the circle
to which the test closed string state should be glued. The properties of the
solution are encoded by insertions on cuts, summing over all possible
amounts and locations of the cuts. Here, we plot two such cuts
(green). The factors of $[B_R,\Psi]$ are glued into this cut by identifying
the left and right parts of $\Psi$ with the two sides of the open cut. This
ensures that the string mid-point touches the red line. Note, that while
``most of the $B_R$ line integral'' can be freely deformed, the point where
it touches the closed string (red line) is fixed. In contrast to that, $B_R$
can be deformed along the open string boundary (blue line) in light of the
doubling trick.}
}

The simplest way to show that the boundary state for Erler's solution is
identical to that of Schnabl's solution is to show that the second term
in the r.h.s of~(\ref{ErlerSol}) does not contribute to $[B_R,\Psi]$.
Now, we face a difficulty, since the exact form of the gluing
should be defined in order to perform explicit calculations, e.g., the
functional form of the integrand of $B_R$ changes upon crossing the lines
where $\Psi$ is glued and this change can be different in the right and left
sides. Hence, treating $B_R$ as a genuine contour integral is too naive.
Luckily, this issue was resolved in~\cite{Kiermaier:2008qu}, by giving
explicit expressions for the case of a Schnabl gauge propagator. Then, a
conformal transformation to a coordinate system similar to the
cylinder coordinates is performed. In this coordinate system the segments
between the cuts are mapped to ``slanted wedges'' and this ``slanting''
of the wedges introduces a hidden boundary at infinity, which is the closed
string boundary.
Explicit calculations can be performed, at least for wedge-state-based
solutions, as we have here.
In the case at hand, we can use eq.~(6.23) of~\cite{Kiermaier:2008qu},
in order to replace the $B_R$ in $[B_R,\tilde \Psi]$ by a sum of a genuine
contour integral and the standard $B$ line integral,
\begin{equation}
[B_R,\tilde \Psi]\rightarrow k_1\oint dz \big((z-z_1) b(z)
  B\gamma^2(z_2)\big)+k_2 B\gamma^2(z_2)B\,.
\end{equation}
Here, $k_{1,2}$ and $z_{1,2}$ are known constants, which are of no
importance for us.
The contour integral can be closed, since no explicit
$c$ insertions are present\footnote{The implicit ghosts in the Virasoro
generators were already taken care of in defining the gluing.}, while the
second term vanishes since the $B$ line integrals can be moved until they
annihilate each other.
It follows that, in each cut, Erler's solution contributes the same as
Schnabl's solution. It follows that the boundary state associated with
Erler's solution is identically zero as stated.
We conclude that this solution indeed represents the closed string vacuum.

\section{Conclusions}
\label{sec:conc}

The classical equivalence of the various cubic superstring field theories
is an important step towards a credible cubic superstring field theory.
However, the fact that our proof is classical implies that off-shell (in the
sense of string field theory) string fields, might have their action depend
on $Y_{-2}$.
Also, there in no sense in which we could have defined it quantum
mechanically, due to the problems with defining the Ramond sector.
Indeed, finding a consistent definition for the Ramond sector seems as the
predominant obstacle towards a sensible cubic theory.
After this problem is resolved, one would have to address the quantum
equivalence, e.g., study whether (loop) amplitudes give the same results
regardless of $Y_{-2}$.

We proved that Erler's solution corresponds to the closed string vacuum,
regardless of the question of existence of a tachyon field. This attribute
caused the initial mistrust of this solution, since previous study of
non-perturbative vacua in string field theory focused on tachyon
condensation. A related question is whether the original and final states
are continuously connected.
It was already suggested that in some sense one can think of the closed
string vacuum as being continuously connected to the perturbative
one~\cite{Fuchs:2008zx}\footnote{Note that, at any rate, the identification
of the closed string vacuum as the end point of tachyon condensation is
not trivial. The marginal deformation that corresponds to tachyon
condensation describes the absence of the original D-brane together with
the radiation emitted during the condensation process~\cite{Sen:2002in}.
In particular, the energy of the solution describing the marginal
deformation is always zero. There are also technical problems with this
identification, such as wide oscillations~\cite{Moeller:2002vx,
Coletti:2005zj,Schnabl:2007az,Kiermaier:2007ba}. Nonetheless, there is a
sense in which the closed string vacuum can be identified with the endpoint
of tachyon condensation~\cite{Ellwood:2007xr}.}.
Anyhow, we find it very encouraging that string field theory is capable of
describing this case as well. It is very desirable to unveil the full realm
of use of string field theory. In particular, in is interesting to find out
whether it is capable of describing multi D-brane solutions when the original
BCFT is that of a single D-brane.

The construction of the boundary state can be naturally generalized to the
case of a non-BPS D-brane~\cite{Berkovits:2000hf,Arefeva:2002mb}. Here,
the NS+ string field $\Psi_+$ is tensored with the ``internal Chan-Paton''
factor $\sigma_3$, while the NS$-$ string field $\Psi_-$ is tensored with
$i\sigma_2$.
The integral in the action includes now also a normalized trace over the
internal Chan-Paton space. 
It is clear that the internal Chan-Paton space that appear in $\Psi$
should be eliminated in order to obtain the boundary state in the correct
space. The natural way to do that is to include a normalized trace in the
definition of the boundary state. Then, in order to obtain the previous
results for $\Psi$ that fully resides in the NS+ space, the $\sigma_3$
factor of $\Psi_+$ should be eliminated as well. Since $\Psi$ enters
the construction in the combination $[B_R,\Psi]$, what we need is to
append $B_R$ with a factor of $\sigma_3$. It seems that these slight
modifications are all that is needed.
Erler's solution was generalized (on the non-BPS D-brane)
in~\cite{Aref'eva:2008ad,Fuchs:2008zx} to a one-parameter family of
solutions. It was claimed in~\cite{Fuchs:2008zx}
that these solutions are all gauge equivalent.
It would be interesting to verify explicitly that these solutions also
correspond to a vanishing boundary state (or a gauge transformation thereof).

\section*{Acknowledgments}

I would like to thank Udi Fuchs, Barton Zwiebach and especially Michael
Kiermaier for many discussions on the issues covered in this work.
I would also like to thank Michael Kiermaier for commenting on the
manuscript.
This work is supported by the U.S. Department of Energy (D.O.E.) under
cooperative research agreement DE-FG0205ER41360.

My research is supported by an Outgoing International Marie Curie
Fellowship of the European Community. The views presented in this work are
those of the author and do not necessarily reflect those of the European
Community.

\bibliography{bib}

\end{document}